# Novel microwave near-field sensors for material characterization, biology, and nanotechnology

R. Joffe, E.O. Kamenetskii, and R. Shavit

Microwave Magnetic Laboratory
Department of Electrical and Computer Engineering
Ben Gurion University of the Negev, Beer Sheva, Israel

**Abstract**

The wide range of interesting electromagnetic behavior of contemporary materials requires that experimentalists working in this field master many diverse measurement techniques and have a broad understanding of condensed matter physics and biophysics. Measurement of the electromagnetic response of materials at microwave frequencies is important for both fundamental and practical reasons. In this paper, we propose a novel near-field microwave sensor with application to material characterization, biology, and nanotechnology. The sensor is based on a subwavelength ferrite-disk resonator with magnetic-dipolar-mode (MDM) oscillations. Strong energy concentration and unique topological structures of the near fields originated from the MDM resonators allow effective measuring material parameters in microwaves, both for ordinary structures and objects with chiral properties.

## I. Introduction

Implementation of imaging in microwave frequencies gives the opportunity for electrodynamics experiments with natural materials and artificial structures. In many microwave sensing devices, resonant structures are essential elements because they allow localization of high field areas. They are very efficient in the frequency band for which they were designed, since the signal-to-noise ratio in a resonator structure increases with resonator quality factor Q. This increase in sensitivity and field strength is accompanied by a narrower frequency band, with the drop in amplitude depending on Q, which results from the shift in resonant frequency with different environments. Consequently, resonance frequency and amplitude tracking are employed. Typical devices for imaging in microwaves use conventional resonant structures. There are, for example, narrow resonant slots in a rectangular hollow waveguide, strip- and microstrip-line resonators, coaxial-cavity and coaxial-line resonators. A detailed description of these structures can be found in review articles [1, 2]. Recently, open planar *LC* resonators have been suggested [3] for microwave sensing.

For subwavelength characterization of microwave material parameters, special metallic probes are mostly used. The near fields of such metallic probes are well known evanescent-mode fields [1, 2]. There is a trade-off between the quality factor of a resonator and the coupling between an object and a tip. To improve the resolution and bring it well below the free-space electromagnetic wave length, different ways for optimization of the scanner tip structures had been suggested. It becomes clear that new perfect lenses that can focus beyond the diffraction limit could revolutionize near-field microwave microscopy. In the present study, we propose a novel microwave near-field sensor with application to material characterization, biology, and nanotechnology. This sensor is realized based on a small ferrite-disk resonator with magnetic-dipolar-mode (MDM) oscillations [4 – 7]. The wavelength of the MDM oscillations in ferrite resonators is two-four orders of the magnitude less than the free-space electromagnetic-wave wavelength at the same microwave frequency [8]. Application of these properties in near-field microwave microscopy allows achieving submicron resolution much easier than in the existing microwave microscopes with standard resonant structures [1, 2]. Moreover, since a MDM ferrite resonator is a multiresonance structure, a complete-set mode spectrum of MDM oscillations can be used to get a complete Fourier image (in the frequency or $\vec{k}$-space domain) in a localized region of a sample [4 – 7, 9].

Due to the growing interaction between biological sciences and electrical engineering disciplines, effective sensing and monitoring of biological samples becomes an important subject. This, especially, concerns the near-field microwave microscopy of chemical and biological structures [3, 10 – 13]. One of the significant questions, both for fundamental studies and applications, is biophysical modeling of microwave-induced nonthermal biological effects [14, 15]. Despite the fact that reports of nonthermal microwave effects date back to the 1970s, there is a great deal of renewed interest. The rapid rate of adoption of mobile phones and mobile wireless communications into society has resulted in public concern about the health hazards of microwave fields emitted by such devices. Direct detection of biological structures in microwave frequencies and understanding of the molecular mechanisms of nonthermal microwave effects is a problem of a great importance. Nowadays, however, microwave biosensing is represented by standard microwave techniques for measuring dielectric and conductivity properties of materials [3, 10 – 13]. Proper correlation of these material parameters with structural characteristics of chemical and biological objects in microwaves appears as a serious problem.

Compared to microwave biosensing, optical biosensing is represented in a larger variety of effective tools. Examining the structure of proteins by circular dichroism is an effective and well known optical technique [16]. Another technique is based on the resonant subwavelength interactions between plasmon (or electrostatic) oscillations of metal nanoparticles and electromagnetic fields. Localized surface plasmon resonance spectroscopy of metallic nanoparticles is a powerful tool for chemical and biological optical sensing [17]. For biomedical diagnostics and pathogen detection, special plasmonic structures with left- and right-handed optical superchiral fields have been recently proposed. These structures effectively interact with large biomolecules, in particular, and chiral materials in general [18].

Chirality is of fundamental interest for chemistry and biology not only in optics, but, certainly, also in microwaves. However, the near-field patterns of nowadays microwave sensors do not have symmetry breakings and so cannot be used for microwave characterization of chemical and biological objects with chiral properties as well as chiral metamaterials. Can one use the main ideas and results of the optical subwavelength photonics to realize microwave structures with subwavelength confinement and chirality of the near fields? Since resonance frequencies of electrostatic (plasmon) oscillations in small particles are very far from microwave frequencies, an answer to this question is negative. It becomes sufficiently apparent that in microwaves, the problem of effective characterization of chemical and biological objects will be solved when one develops special sensing devices with microwave chiral probing fields. Recently, it has been shown that the MDM ferrite particles may create microwave superchiral fields with strong subwavelength localization of electromagnetic energy. There are the particles with magnetostatic (MS) (or MS-magnon) oscillations, but not with electrostatic (plasmon) oscillations [19 – 22].

In this paper, we propose a novel near-field microwave sensor with application to material characterization, biology, and nanotechnology. The sensor is based on a small ferrite-disk resonator. The MDM oscillations in this resonator are characterized by time and space symmetry breakings. Development of such subwavelength sensors for direct microwave characterization of microscopic material structure with application to biology and nanotechnology is a subject of high importance. In microwave near-field sensing with superchiral fields, we have a completely new mechanism of the material-field interaction. "New truths become evident when new tools become available"; these words of Nobel Laureate Rosalyn Yalow may concern also the proposed near-field microwave sensors.

**II. Microwave superchiral fields**

The near fields originated from a normally magnetized ferrite disk with MDM oscillations have intrinsic chiral topology. They are characterized by strong subwavelength localization of energy and vortices of the power-flow density. In the vicinity of a ferrite disk, one can observe the power-flow whirlpool. Fig. 1 illustrates the field structure at the resonance frequency of MDM oscillation. No such confinement and topology of the fields are observed at nonresonance frequencies. An electric field inside a ferrite disk has both orbital and spin angular momentums [23]. This results in helical-mode resonances [7, 20]. When an electrically polarized dielectric sample is placed above a ferrite disk, every separate dipole in a sample will precess around its own axis. For all the precessing dipoles, there is an orbital phase running [see Fig. 2].

The mechanical torque exerted on a given electric dipole is defined as a cross product of the MDM electric field $\vec{E}$ and the electric moment of the dipole $\vec{p}$ [21, 22]:

$$\vec{N} = \vec{p} \times \vec{E}. \tag{1}$$

The dipole $\vec{p}$ appears because of the electric polarization of a dielectric by the RF electric field of a microwave system. The torque exerting on the electric polarization due to the MDM electric field should be equal to reaction torque exerting on the magnetization in a ferrite disk. Because of this reaction torque, the precessing magnetic moment density of the ferromagnet will be under additional mechanical rotation at a certain frequency $\Omega$. For the magnetic moment density of the ferromagnet, $\vec{M}$, the motion equation acquires the following form [21, 22]:

$$\frac{d\vec{M}}{dt} = -\gamma \, \vec{M} \times \left( \vec{H} - \frac{\Omega}{\gamma} \right), \tag{2}$$

The frequency $\Omega$ is defined based on both, spin and orbital, momentums of the fields of MDM oscillations. One can see that at dielectric loadings, the magnetization motion in a ferrite disk is characterized by an effective magnetic field

$$\vec{H}_{eff} = \vec{H} - \frac{\Omega}{\gamma}. \tag{3}$$

So, the Larmor frequency of a ferrite structure with a dielectric loading should be lower than such a frequency in an unloaded ferrite disk.

The near fields originated from MDM ferrite particles are characterized by subwavelength confinement of energy and chirality properties. In optics, a parameter characterizing superchiral fields is a so-called optical chirality density [18]. For quantitative characterization of microwave superchiral fields created by MDM ferrite particles, we use the helicity density factor *F*. For time-harmonic fields, parameter *F* is expressed as [21, 22]:

$$F = \frac{\varepsilon}{4} \text{Im} \left\{ \vec{E} \cdot \left( \vec{\nabla} \times \vec{E} \right)^* \right\}. \qquad (4)$$

From this equation one can establish the essential property of local microwave fields which possess chiral symmetry: the components of complex vectors $\vec{E}$ and $\vec{\nabla} \times \vec{E}$ must be parallel and phase shifted giving rise a nonzero imaginary component to their dot product. It is worth noting that a sign of the helicity factor of a microwave superchiral field originated from a MS-magnon mode depends on a direction of a bias magnetic field $\vec{H}_0$. This is one of the main points in our work.

**III. Material characterization**

The MDM spectral characteristics in thin ferrite disks are well observed in microwave experiments [24 – 27], can be studied analytically [4 – 7], and are well illustrated by numerical results based on the HFSS Ansoft solver [19, 20, 23]. Fig. 3 shows the numerically simulated MDM spectral characteristic for a thin ferrite disk placed in a $TE_{10}$-mode rectangular waveguide. An insert in the figure shows geometry of the structure. The yttrium iron garnet (YIG) disk has a diameter of $D = 3$ mm and a thickness of $t = 0.05$ mm. The saturation magnetization of a ferrite is $4\pi M_s = 1880$ G. The disk is normally magnetized by a bias magnetic field $H_0 = 4900$ Oe. For better understanding the field structures, in numerical studies we use a ferrite disk with very small losses: the linewidth of a ferrite is $\Delta H = 0.1$ Oe. The waveguide walls are made of a perfect electric conductor (PEC). In the spectrum in Fig. 3 one sees the module of the reflection (the $S_{11}$ scattering-matrix parameter) coefficient. The resonance modes are designated in succession by numbers $n = 1, 2, 3…$ It is evident that starting from the second mode, the coupled states of the electromagnetic fields with MDM vortices are split-resonance states. The properties of these coalescent resonances, denoted in Fig. 3 by single and double primes, were analyzed in details in Ref. [20]. The near fields at the MDM resonances are characterized by the helicity properties. Fig. 4 gives evidence for distinct parameters of helicity density of the fields, calculated based on Eq. (4).

Topological effects appearing in the matter-field interaction for the microwave superchiral fields open a perspective for unique near-field characterization of material parameters. Due to strong energy localization and the field helicity one has novel tools for effective measuring material parameters in microwaves. As an example of such measuring, we use a structure composed by a MDM ferrite disk and two dielectric cylinders loading symmetrically a ferrite disk. Fig. 5 (a) shows this structure placed inside a $TE_{10}$-mode rectangular waveguide. The dielectric cylinders (every cylinder is with the diameter of 3 mm and the height of 2 mm) are electrically polarized by the RF electric field of the $TE_{10}$ mode propagating in a waveguide. The dielectric loadings do not destroy the entire MDM spectrum, but cause, however, the frequency shifts of the resonance peaks. The frequency characteristics of a module of a reflection coefficient for the 1st MDM resonance at different dielectric parameters of the cylinders are shown in Fig. 5 (b). One can see that at dielectric loadings, the 1st MDM resonance appears as coalescent resonances (the resonances 1′ and 1″ ). One of the main features of the frequency characteristics of a structure with the symmetrical dielectric loadings, is the fact that the resonances of the 1st MDM become shifted not only to the lower frequencies, but appear to the left of the Larmor frequency of an unloaded ferrite disk. For a normally magnetized ferrite disk with the pointed above quantities of the bias magnetic field and the saturation magnetization, this Larmor frequency (calculated as $f_H = \frac{1}{2\pi} \gamma H_i$, where $\gamma$ is the gyromagnetic ratio and $H_i$ is the internal DC magnetic field) is equal to $f_H = 8,456$ GHz. When a ferrite disk is without dielectric loadings, the entire spectrum of MDM oscillations is situated to the right of the Larmor frequency $f_H$. Since dielectrics do not destroy the entire MDM spectrum, one can suppose that the Larmor frequency of a structure with dielectric loadings $f_H^{(D)}$ is lower than the Larmor frequency of an unloaded ferrite disk $\left( f_H^{(D)} < f_H \right)$. This statement is well clarified by the above analysis with use of Eqs. (1) – (3). It is worth noting that dielectric loadings change distribution of the helicity density. The helicity-parameter distributions calculated based on Eq. (4) are shown in Fig. 6 for different dielectric constants of loading cylinders. The frequencies correspond to the resonance-peak positions in Fig. 6. As one can see, the dielectric loadings not only reduce the quantity of the helicity factor, but result in strong modification of the near-field structure.

For practical purposes in the microwave characterization of materials, it is more preferable to use an open-access microstrip structure with a ferrite-disk sensor, instead of a closed waveguide structure studied above. Such a microstrip structure is shown in Fig. 7. Fig. 8 represents the frequency characteristic of a module of the transmission (the $S_{21}$ scattering-matrix parameter) coefficient for a microstrip structure with a thin-film ferrite disk. While in a discussed above waveguide structure with an enclosed ferrite disk, the main features of the MDM spectra are evident from the reflection characteristics, in the shown microstrip structure, the most interesting are the transmission characteristics. Classification of the resonances shown in Fig. 8 is made based on analytical studies in Ref. [5]. There are resonances corresponding to MDMs with radial and azimuth variations of the magnetostatic-potential wave functions in a ferrite disk. One can see that the frequencies of the 1$^{st}$ and 2$^{nd}$ radial-variation resonances shown in Fig. 8 are in a good correspondence with the frequencies of the 1$^{st}$ and 2$^{nd}$ resonances shown in Fig. 3 for a waveguide structure. Between the 1$^{st}$ and 2$^{nd}$ resonances of the radial variations one can see the resonance of the azimuth-variation mode. This resonance appears because of the azimuth nonhomogeneity of a microstrip structure. For experimental studies, we use a ferrite disk with the same parameters as pointed above for numerical analyses. The only difference that in an experimental disk the linewidth of a ferrite is $\Delta H = 0.8$ Oe. An experimental microstrip structure is realized on a dielectric substrate (Taconic RF-35, $\varepsilon_r = 3.52$, thickness of 1.52 mm). Characteristic impedance of a microstrip line is 50 Ohm. For dielectric loadings, we used cylinders of commercial microwave dielectric (non magnetic) materials with the dielectric permittivity parameters of $\varepsilon_r = 30$ (K-30; TCI Ceramics Inc) and $\varepsilon_r = 50$ (K-50; TCI Ceramics Inc). A transmission coefficient was measured with use of a network analyzer. With use of a current supply we established a quantity of a normal bias magnetic field $\vec{H}_0$, necessary to get the MDM spectrum in a required frequency range. It is evident that there is a sufficiently good correspondence between the numerical and experimental results of transformation of the MDM spectra due to dielectric loadings. It is necessary to note, however, that instead of a bias magnetic field used in numerical studies ($H_0 = 4900$ Oe), in the experiments we applied lower quantity of a bias magnetic field: $H_0 = 4708$ Oe. Use of such a lower quantity (giving us the same positions of the non-loading-ferrite resonance peaks in the numerical studies and in the experiments) is necessary because of non-homogeneity of an internal DC magnetic field in a real ferrite disk. A more detailed discussion on a role of non-homogeneity of an internal DC magnetic field in the MDM spectral characteristics can be found in Refs. [5, 25].

For effective localization of energy of MDM oscillations at micron and submicron near-field regions, special field concentrators should be used. In particular, there can be a thin metal wire placed on a surface of a ferrite disk. Fig. 9 shows a microstrip MDM sensor with a wire concentrator. A bias magnetic field $\vec{H}_0$ is directed normally to a disk plane. The wire electrode has diameter of 100 um. In a shown structure of a ferrite disk with a wire electrode, the helical waves localized in a ferrite disk are transmitted to the end of a wire electrode. The electric field of a microstrip structure causes a linear displacement of charge when interacting with a short piece of a wire, whereas the magnetic field of a MDM vortex causes a circulation of charge. These two motions combined cause an excitation of an electron in a helical motion, which includes translation and rotation. Fig. 10 illustrates this effect. The electric field distributions on a wire electrode, shown for two different time phases, give evidence for helical waves. Due to such helical waves, at a butt end of a wire one has electric and magnetic fields with mutually parallel components [see Fig. 11 (*a*), (*b*)] and a chiral surface electric current [see Fig. 11 (*c*)]. All this results in appearing of the power-flow-density vortex [see Fig. 12 (*a*)] and nonzero helicity density *F* [see Fig. 12 (*b*)] at the butt end of a wire electrode. Fig. 13 (*a*) shows a microstrip structure with a field concentrator for localized material characterization. Experimental results of the MDM spectra transformations for different dielectric samples are shown in Fig. 13 (*b*).

**IV. Microwave chirality discrimination**

A crucial point on the proposed microwave sensor is the fact that a sign of the helicity factor of a microwave superchiral field originated from a MDM ferrite particle depends on a direction of a bias magnetic field $\vec{H}_0$. Fig. 14 shows distribution of the helicity density of the MDM near fields at two opposite orientations of a normal bias magnetic field. This distribution is obtained for a ferrite disk placed in a microstrip structure. The disk is without a dielectric loading. Because of dependence of a sign of the helicity parameter on orientation of a bias magnetic field, one can propose microwave chirality discrimination using a ferrite-disk sensor loaded by "right" and "left" enantiomeric structures. Fig. 15 illustrates discrimination of enantiomeric structures by the numerically simulated MDM spectra at two opposite orientations of a bias magnetic field $\vec{H}_0$. As a test structure, mimicking an object with chiral properties, a small dielectric sample with symmetry-breaking geometry of surface metallization is used. Investigations are made based on a microstrip structure shown in Fig. 7. For the "right" and "left" structures loading a ferrite disk, there are clearly observed frequency shifts between high-order resonance peaks in the reflection spectra.

For discrimination of enantiomeric structures at localized regions, a microstrip MDM sensor with a wire concentrator can be used. The field properties shown in Fig. 12 give evidence for the symmetry breakings. When one changes oppositely an orientation of a bias magnetic field $\vec{H}_0$, one has an opposite rotation of the power flow and an opposite sign of the helicity density *F* (red colored instead of blue colored). This allows prediction of localized sensing samples with "right" and "left" handedness. As a localized test structures mimicking objects with chiral properties, we used small metallic helices. Fig. 16 shows the setup of the measurement system with a helical test structure for local determination of material chirality. A wire concentrator is placed near a metallic helix without an electric contact with it. Detailed pictures of the sensor with small left- and right-handed helix particles are shown in Fig. 17. In Fig. 18, one can see numerical results of the transmission coefficients for the left-handed and right-handed helical test structure, respectively. The spectral characteristics are obtained for two opposite orientations of a bias magnetic field $\vec{H}_0$. It is worth noting that spectral recognition of enantiomeric structures at opposite orientations of a bias magnetic field shown in Fig. 15 (a) is different from the results of spectral discrimination shown in Fig. 18. While in the former case loading of a ferrite disk by enantiomeric samples leads to frequency shifts in the reflection spectrum, in the last case, one observes amplitude differences in the resonance peaks of the transmission spectrum. A helical test sample shown in Fig 16 very slightly loads a ferrite disk and no frequency transformations in the transmission and reflection spectra are observed in this case.

It is also worth noting that the results in Fig. 18 exhibit very specific symmetry properties. Following these results one observes restoration of an entire transmission spectrum when handedness of a sample is changed together with change of direction of a bias magnetic field. As it was shown in Refs [7, 20, 21], helicity of the near field is originated from double-helix resonances of MDM oscillations in a quasi-2D ferrite disk. For two helical modes, giving a double-helix resonance in a ferrite disk, there is no parity ($\mathcal{P}$) and time-reversal ($\mathcal{T}$) invariance – the $\mathcal{PT}$-invariance. Such $\mathcal{PT}$-symmetry breaking does not guarantee real-eigenvalue spectra. It was shown, however [7], that by virtue of quasi-two dimensionality of the MDM spectral problem, one can reduce solutions from helical to cylindrical coordinates and such a cylindrical-coordinate problem has the $\mathcal{PT}$-invariant solutions. On the other hand, change of handedness of a helical test structure cannot be considered as the space reflection operation. Certainly, electric currents induced in enantiometric structures (metallic helices) by MDM near fields are not $\mathcal{PT}$-invariant quantities. It is evident that specific symmetries (sample handedness ↔ bias magnetic field) of the spectra in Fig. 18 are not related to the $\mathcal{PT}$-symmetry. At the same time, these symmetries highlight unique topological properties of microwave superchiral fields.

**V. Discussion on credibility of the numerical results**

One of the main questions for discussions may concern credibility of the results, obtained based on the classical-electrodynamics HFSS program, for description of the shown non-trivial topological effects originated from the MDM ferrite disks. The properties of the MDM oscillations inside a ferrite disk are well observed numerically due to the ANSOFT HFSS program [19, 20, 23]. From the HFSS studies of microwave structures with thin-film ferrite disks, one has strong regularity of the results. One can see consistent pattern of both the spectral-peak positions and specific topological structure of the fields of the oscillating modes. One observes the same eigenresonances and the same field topology of the resonance eigenstates in different guiding structures [when, for example, a ferrite disk is placed in a rectangular waveguide and when it is placed in a microstrip structure]. There is a very good correspondence of the HFSS results with the analytical and experimental results.

It is worth noting that with use of the HFSS program for our studies we are faced with numerical solutions of a non-integrable (path-dependent) electromagnetic problem. In a case of ferrite inclusions in microwave structures, acting in the proximity of the ferromagnetic resonance, the phase of the electromagnetic wave reflected from a ferrite boundary depends on the direction of the incident wave. For a given orientation of a bias magnetic field, one can distinguish the right-hand and left-hand rays of electromagnetic waves. This fact, arising from peculiar boundary conditions of the fields on the dielectric-ferrite interfaces, leads to the time-reversal symmetry breaking effect in microwave resonators with inserted ferrite samples. A nonreciprocal phase behavior for electromagnetic fields on a surface of a ferrite disk results in the problem of path-dependent numerical integration. For the spectral-problem solutions, the HFSS program, in fact, composes the field structures from interferences of multiple plane EM waves inside and outside a ferrite particle. In such a numerical analysis one obtains the pictures of the field structures (resulting in the real-space integration) based on integration in the *k*-space of the EM fields. In the *k*-space integration, the fields are expanded from very low wavenumbers (free-space EM waves) to very high wavenumbers (the region of MS oscillations). It is also worth noting that inside (and nearly outside) a ferrite disk, the HFSS numerical results can be well applicable for confirmation of the analytically-derived quantum-like models based on the magnetostatic-potential scalar wave function $\psi$. There exist many examples in physics when non-classical effects are well modeled by a combined effect of numerous classical sub-elements.

**VI. Conclusion**

Presently subwavelength characterization of microwave material parameters is considering as a very topical subject. An evident progress in scanning near-field microwave microscopy allows make precise measurements of different material structures on submicron scales. Among these material structures there are, in particular, biological systems. One can state, however, that nowadays for biological structure characterizations, optical techniques are considered as the most effective. Together with the well known techniques of circular dichroism, there are different plasmonic methods, including recently developed optical measurements based on so-called superchiral near fields. All these techniques are not applicable to microwaves.

The problem of effective characterization of chemical and biological objects in microwaves can be solved when one develops special sensing devices with microwave chiral probing fields. In this paper, we showed that small ferrite-disk resonators with magnetic-dipolar-mode (MDM) oscillations may create microwave superchiral fields with strong subwavelength localization of electromagnetic energy. Based on such properties of the fields, we propose a novel near-field microwave sensor with application to material characterization, biology, and nanotechnology. With use of these sensors one opens unique perspective for effective measuring material parameters in microwaves, both for ordinary structures and objects with chiral properties.

**Figure captions**

Fig. 1. (*a*) Intensity and geometry of the power-flow distribution: strong subwavelength localization of energy and vortex behavior of microwave near fields. (*b*) Schematic picture of the power-flow whirlpool in the vicinity of a ferrite disk.

Fig. 2. An electric field inside a ferrite disk has both orbital and spin angular momentums. When an electrically polarized dielectric sample is placed above a ferrite disk, electric dipoles in a dielectric sample precess and accomplish an orbital geometric-phase rotation. A bias magnetic field $\vec{H}_0$ is directed normally to a disk plane. For an opposite direction of $\vec{H}_0$, one has an opposite rotation of an electric field and an opposite direction of precession of electric dipoles.

Fig. 3. Frequency characteristics of a module of the reflection coefficient for a rectangular waveguide with an enclosed thin-film ferrite disk. The resonance modes are designated in succession by numbers $n = 1, 2, 3…$ The coalescent resonances are denoted by single and double primes. An insert shows geometry of a structure.

Fig. 4. The near-field helicity parameters. (*a*) Near-field helicity for the 1$^{st}$ MDM. (*b*) Near-field helicity for the 2$^{nd}$ (the resonance 2"). (*c*) Absence of the near-field helicity for non-resonance frequencies.

Fig. 5. (a) A sample of a ferrite disk with two loading dielectric cylinders placed inside a $TE_{10}$-mode rectangular waveguide. (b) Frequency characteristics of a module of the reflection coefficient for the 1$^{st}$ MDM at different parameters of a symmetrical dielectric loading. Frequency $f_H = 8,456$ GHz is the Larmor frequency of an unloaded ferrite disk.

Fig. 6. Numerically calculated helicity-parameter distributions for the 1$^{st}$ MDM at different dielectric constants of loading cylinders. The distributions are shown on the cross-section plane which passes through the diameter and the axis of the ferrite disk.

Fig. 7. A microwave microstrip structure (sensor) with a MDM ferrite disk and a sample under investigation.

Fig. 8. Transformation of the MDM spectrum due to a dielectric loading in a microstrip structure. (*a*) Numerical results; (*b*) Experimental results.

Fig 9. A sensor with a wire concentrator localized material characterization. (*a*) Geometry of a microstrip structure; (*b*) A magnified picture of a MDM ferrite disk with a wire electrode.

Fig. 10. Evidence for helical waves. The electric field distributions on a wire electrode for different time phases. (*a*) Time phase $\omega t_1$; (*b*) Time phase $\omega t_2$.

Fig 11. The fields and currents at a butt end of a wire electrode. (*a*) An electric field; (*b*) A magnetic field; (*c*) A chiral surface electric current.

Fig. 12. Field structure on a butt end of a wire concentrator: (*a*) the power-flow-density vortex and (*b*) the helicity density *F*. When one changes oppositely an orientation of a bias magnetic field $\vec{H}_0$, one has an opposite rotation of the power flow and an opposite sign of the helicity density *F* (red colored instead of blue colored).

Fig. 13. A microstrip structure for localized material characterization in a dielectric sample. (*a*) Geometry of a structure; (*b*) MDM spectra of the reflection coefficient at a dielectric loading (experimental results).

Fig 14. Helicity density *F* of the MDM near fields at two orientations of a normal bias magnetic field $\vec{H}_0$. A ferrite disk is without a dielectric loading and is placed in a microstrip structure shown in Fig. 7.

Fig. 15. (*a*) Discrimination of an enantiomeric structure by the MDM spectra at two opposite orientations of a bias magnetic field $\vec{H}_0$. (*b*) Test structure, mimicking an object with chiral properties: a small dielectric sample with symmetry-breaking geometry of surface metallization.

Fig. 16. Setup of the measurement system with a helical test structure for local determination of material chirality.

Fig. 17. Sensor with small helix particles. (*a*) left-handed helix particle; (*b*) right-handed helix particle.

Fig. 18. Transmission coefficients for small helix particles. (*a*) left-handed helix particle; (*b*) right-handed helix particle.

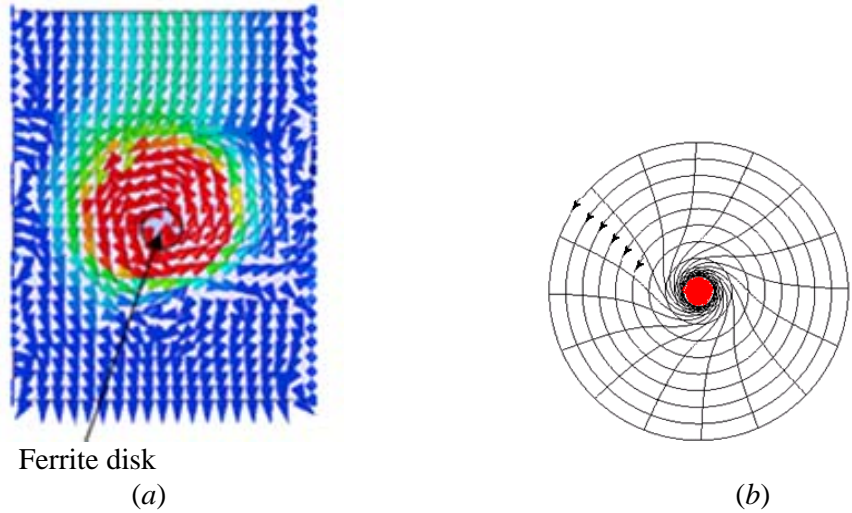

Ferrite disk
(*a*)                                                           (*b*)

Fig. 1. (*a*) Intensity and geometry of the power-flow distribution: strong subwavelength localization of energy and vortex behavior of microwave near fields. (*b*) Schematic picture of the power-flow whirlpool in the vicinity of a ferrite disk.

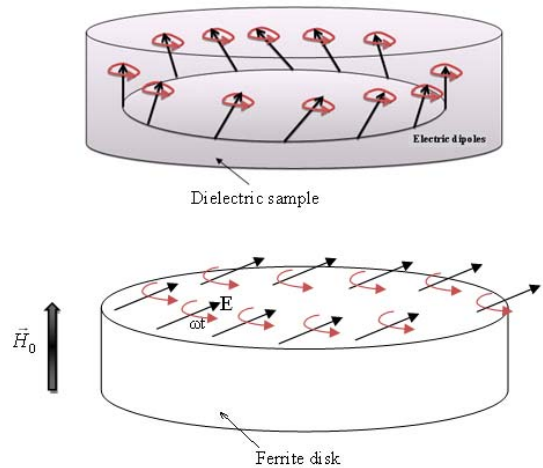

Fig. 2. An electric field inside a ferrite disk has both orbital and spin angular momentums. When an electrically polarized dielectric sample is placed above a ferrite disk, electric dipoles in a dielectric sample precess and accomplish an orbital geometric-phase rotation. A bias magnetic field $\vec{H}_0$ is directed normally to a disk plane. For an opposite direction of $\vec{H}_0$, one has an opposite rotation of an electric field and an opposite direction of precession of electric dipoles.

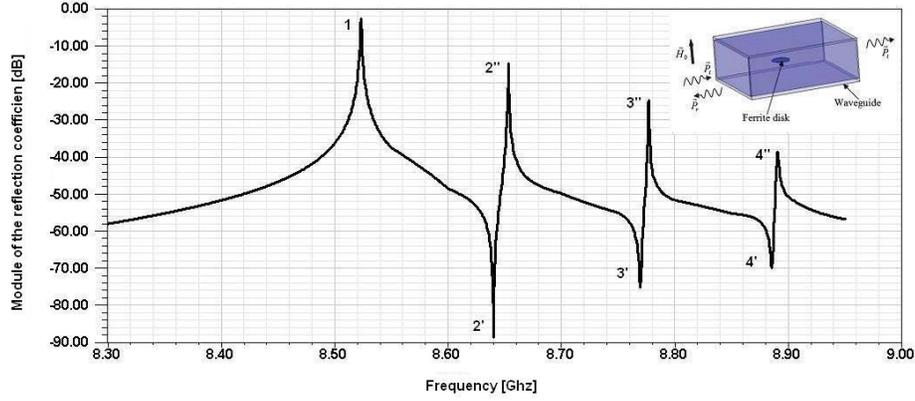

Fig. 3. Frequency characteristics of a module of the reflection coefficient for a rectangular waveguide with an enclosed thin-film ferrite disk. The resonance modes are designated in succession by numbers $n = 1, 2, 3…$ The coalescent resonances are denoted by single and double primes. An insert shows geometry of a structure.

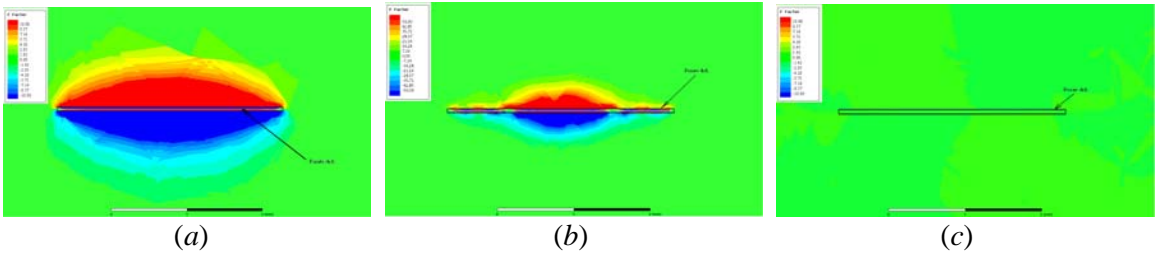

Fig. 4. The near-field helicity parameters. (*a*) Near-field helicity for the 1$^{st}$ MDM. (*b*) Near-field helicity for the 2$^{nd}$ (the resonance 2''). (*c*) Absence of the near-field helicity for non-resonance frequencies.

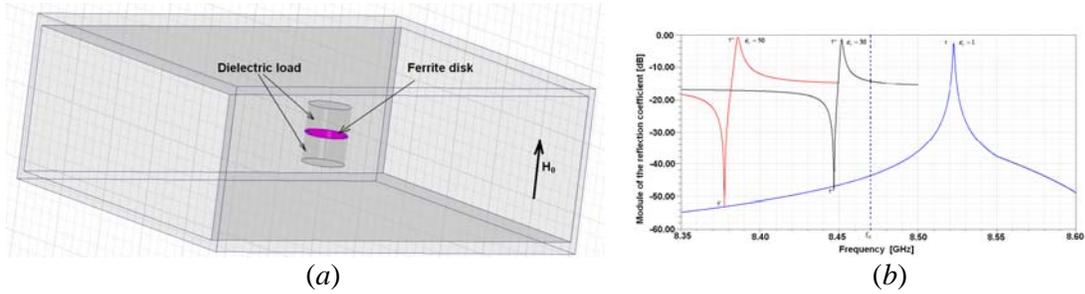

Fig. 5. (*a*) A sample of a ferrite disk with two loading dielectric cylinders placed inside a $TE_{10}$-mode rectangular waveguide. (*b*) Frequency characteristics of a module of the reflection coefficient for the 1$^{st}$ MDM at different parameters of a symmetrical dielectric loading. Frequency $f_H = 8,456$ GHz is the Larmor frequency of an unloaded ferrite disk.

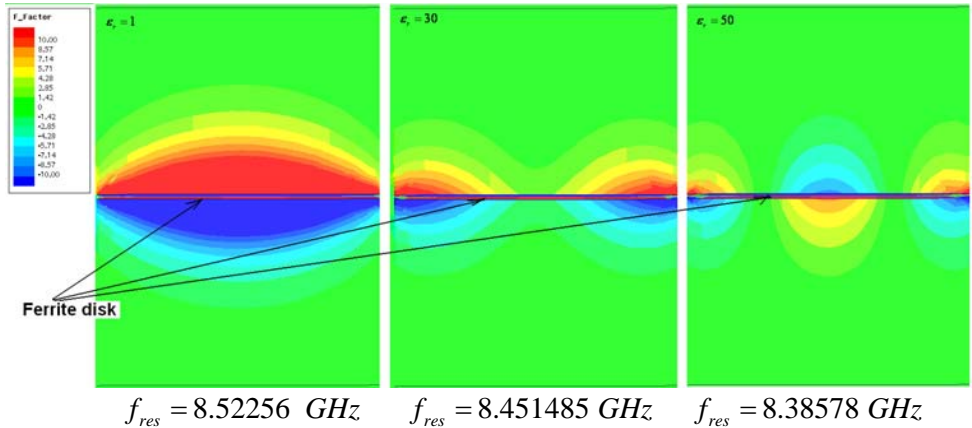

$f_{res} = 8.52256\ GHz \qquad f_{res} = 8.451485\ GHz \qquad f_{res} = 8.38578\ GHz$

Fig. 6. Numerically calculated helicity-parameter distributions for the 1$^{st}$ MDM at different dielectric constants of loading cylinders. The distributions are shown on the cross-section plane which passes through the diameter and the axis of the ferrite disk.

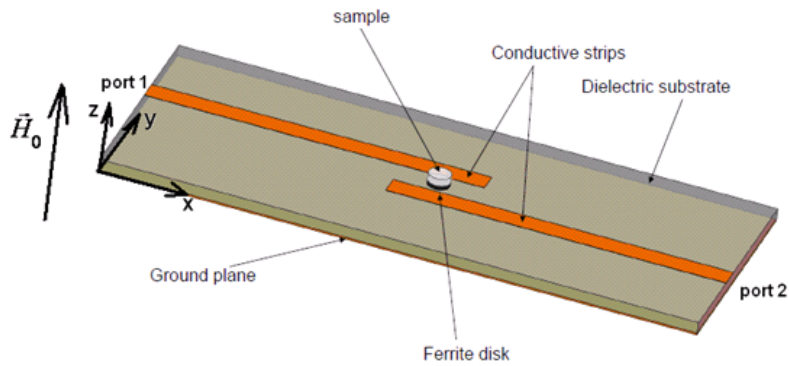

Fig. 7. A microwave microstrip structure (sensor) with a MDM ferrite disk and a sample under investigation.

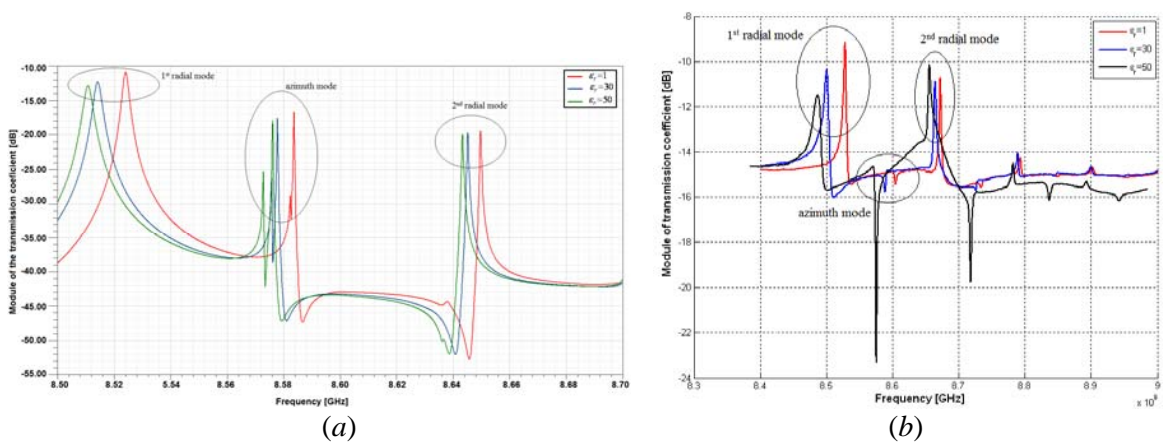

(*a*)            (*b*)

Fig. 8. Transformation of the MDM spectrum due to a dielectric loading in a microstrip structure. (*a*) Numerical results; (*b*) Experimental results.

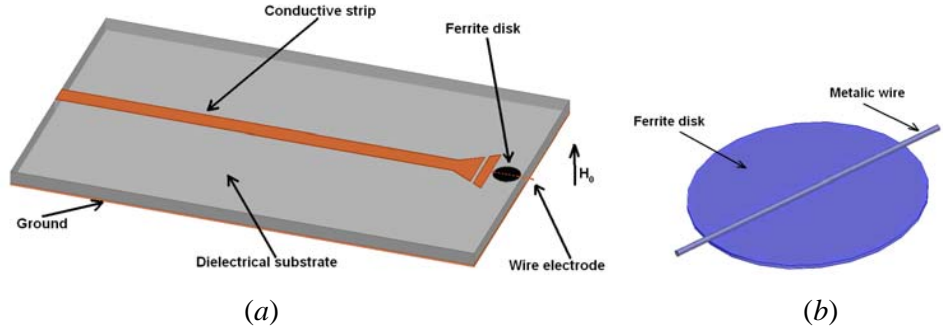

(*a*)                      (*b*)

Fig. 9. A sensor with a wire concentrator for localized material characterization. (*a*) Geometry of a microstrip structure; (*b*) A magnified picture picture of a MDM ferrite disk with a wire electrode.

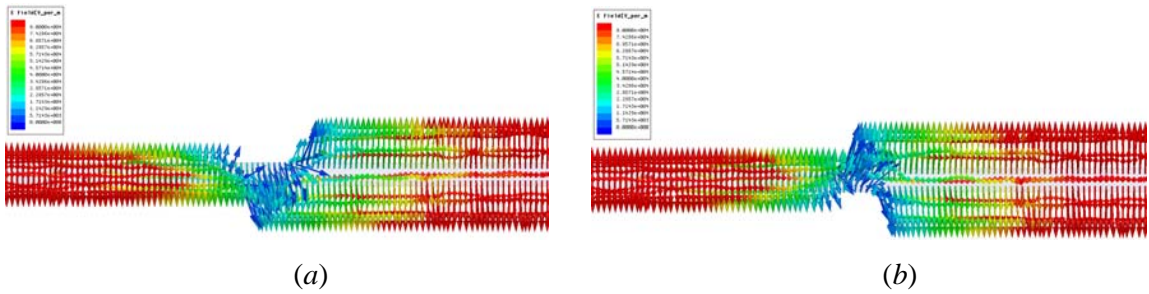

(*a*)                      (*b*)

Fig. 10. Evidence for helical waves. The electric field distributions on a wire electrode for different time phases. (*a*) Time phase $\omega t_1$; (*b*) Time phase $\omega t_2$.

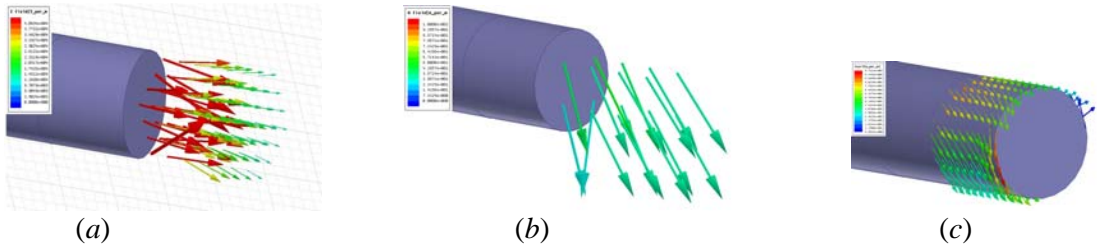

(*a*)            (*b*)            (*c*)

Fig 11. The fields and currents at a butt end of a wire electrode. (*a*) An electric field; (*b*) A magnetic field; (*c*) A chiral surface electric current.

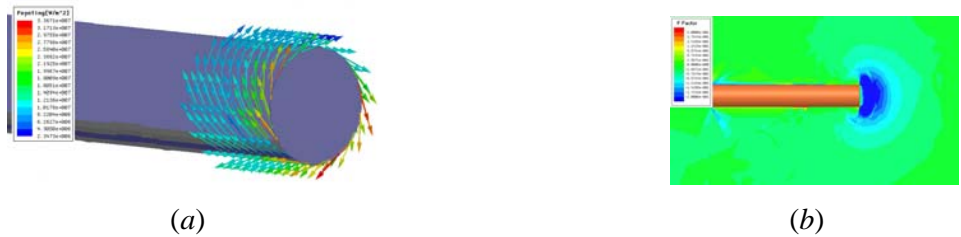

(*a*)                      (*b*)

Fig. 12. Field structure on a butt end of a wire concentrator: (*a*) the power-flow-density vortex and (*b*) the helicity density *F*. When one changes oppositely an orientation of a bias magnetic

field $\vec{H}_0$, one has an opposite rotation of the power flow and an opposite sign of the helicity density *F* (red colored instead of blue colored).

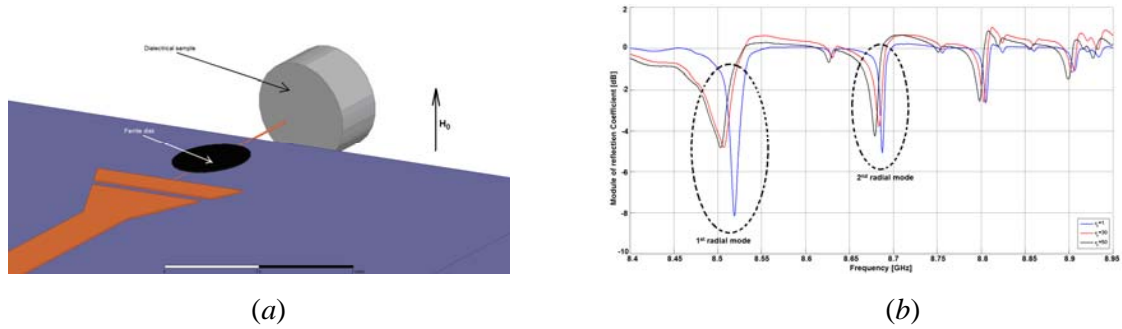

(*a*)                                                    (*b*)

Fig. 13. A microstrip structure for localized material characterization in a dielectric sample. (*a*) Geometry of a structure; (*b*) MDM spectra of the reflection coefficient at a dielectric loading (experimental results).

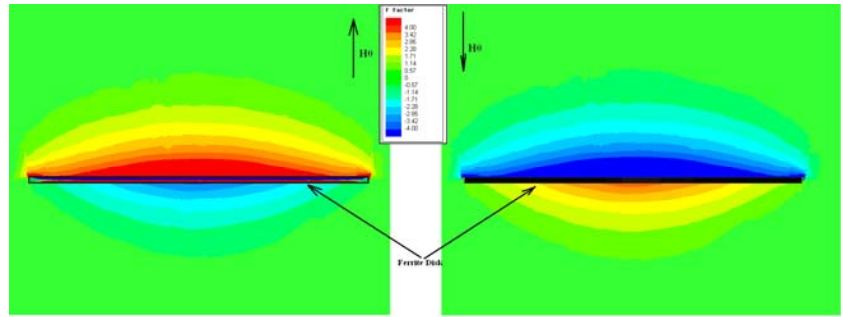

Fig 14. Helicity density *F* of the MDM near fields at two orientations of a normal bias magnetic field $\vec{H}_0$. A ferrite disk is without a dielectric loading and is placed in a microstrip structure shown in Fig. 7.

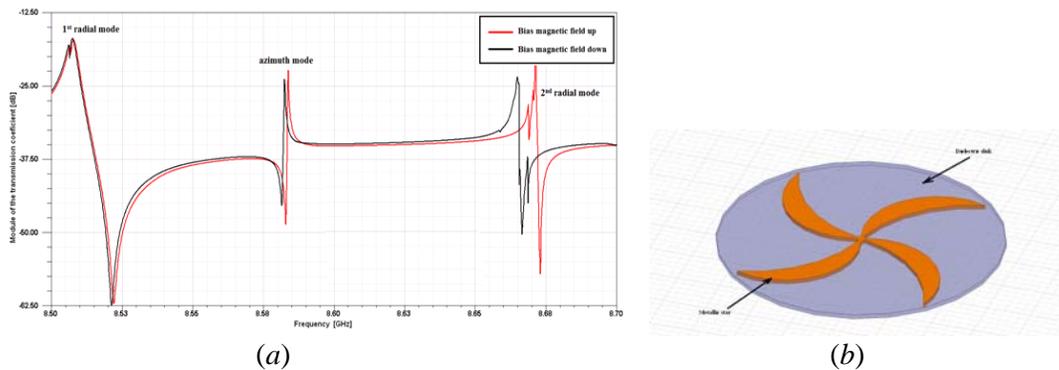

(*a*)                                                    (*b*)

Fig. 15. (*a*) Discrimination of an enantiomeric structure by the MDM spectra at two opposite orientations of a bias magnetic field $\vec{H}_0$. (*b*) Test structure, mimicking an object with chiral properties: a small dielectric sample with symmetry-breaking geometry of surface metallization.

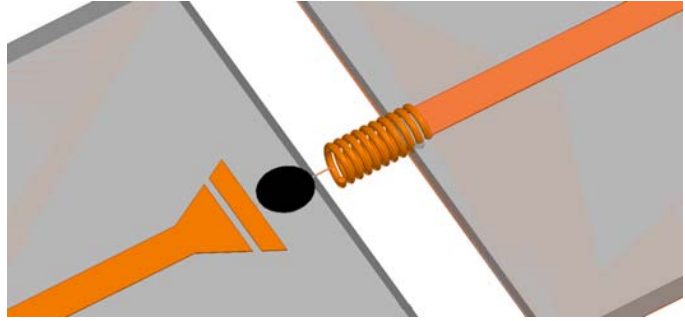

Fig. 16. Setup of the measurement system with a helical test structure for local determination of material chirality.

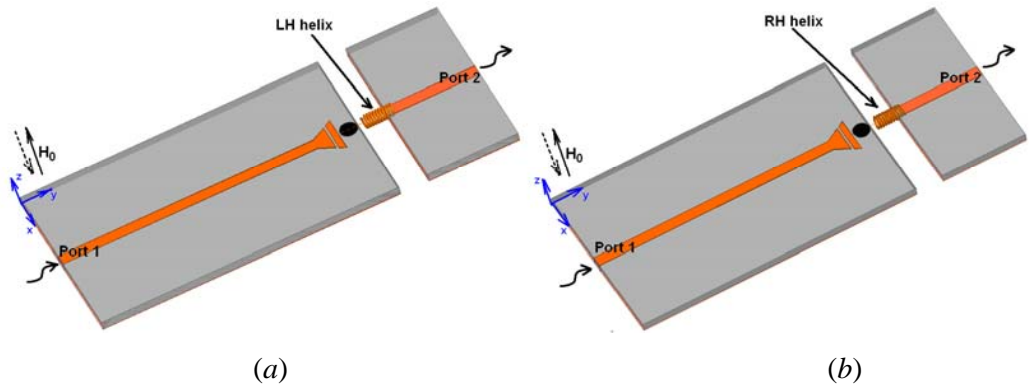

(*a*)            (*b*)

Fig. 17. Sensor with small helix particles. (*a*) left-handed helix particle; (*b*) right-handed helix particle.

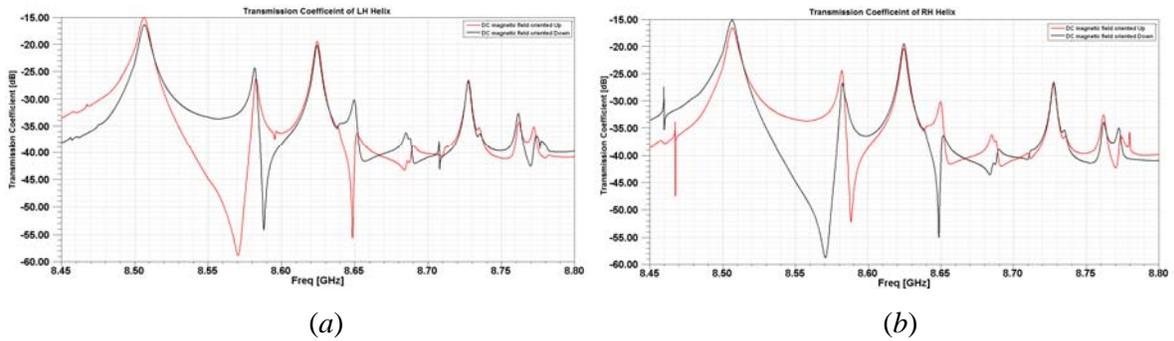

(*a*)            (*b*)

Fig. 18. Transmission coefficients for small helix particles. (*a*) left-handed helix particle; (*b*) right-handed helix particle.

16